\documentclass[ reprint, onecolumn, superscriptaddress, floatfix, amsmath,amssymb, aps ]{revtex4-2}
\usepackage[utf8]{inputenc}
\usepackage[letterpaper,top=2cm,bottom=2cm,left=3cm,right=3cm,marginparwidth=1.75cm]{geometry}

\usepackage{amsmath, amsfonts,amssymb}
\usepackage{physics}
\usepackage{graphicx}
\usepackage[colorlinks=true, allcolors=blue]{hyperref}
\usepackage{dsfont}
\usepackage{hyperref}
\usepackage{mathtools}
\usepackage{braket}
\usepackage{svg}
\usepackage{siunitx}
\usepackage{soul}

\newcommand{\vect}[1]{\mathbf{#1}}
\newcommand{\xv}{\vect{x}}
\newcommand{\mfs}{\mathrm{MFS}}
\renewcommand{\eqref}[1]{Eq.~(\ref{eq:#1})}

\def\XXint#1#2#3{{\setbox0=\hbox{$#1{#2#3}{\int}$}
\vcenter{\hbox{$#2#3$}}\kern-.5\wd0}}

\begin{document}

\title{Multimode grating couplers via foundry-compliant inverse design}

\author{Hao Li}
\affiliation{Department of Electrical and Computer Engineering, Yale University, New Haven, Connecticut
06511, USA}
\affiliation{Energy Sciences Institute, Yale University, New Haven, Connecticut 06511, USA}
\author{Nazar Pyvovar}
\affiliation{Energy Sciences Institute, Yale University, New Haven, Connecticut 06511, USA}
\affiliation{Department of Applied Physics, Yale University, New Haven, Connecticut 06511, USA}
\author{Zhaowei Dai}
\affiliation{Department of Electrical and Computer Engineering, Yale University, New Haven, Connecticut
06511, USA}
\affiliation{Energy Sciences Institute, Yale University, New Haven, Connecticut 06511, USA}
\author{Owen D. Miller}
\email[Corresponding author: ]{owen.miller@yale.edu}
\affiliation{Department of Electrical and Computer Engineering, Yale University, New Haven, Connecticut
06511, USA}
\affiliation{Energy Sciences Institute, Yale University, New Haven, Connecticut 06511, USA}
\affiliation{Department of Applied Physics, Yale University, New Haven, Connecticut 06511, USA}
\date{\today}

\begin{abstract}
We apply a systematic inverse design approach to discover foundry-compliant, multilayer grating couplers that can efficiently couple a number of independent waves from free space to on-chip propagating modes. For visible- and near-infrared couplers, we find that minimum feature sizes are by far the most important constraint to tailor the design algorithms around. If, additionally, one forces the optimization to be robust to over- and under-etch errors, the resulting designs exhibit stable optimal efficiencies in the presence of other imperfections (critical dimension variations, overlay mismatch, and sidewall angle variation). The foundry-compliant designs exhibit moderate efficiency penalties as feature sizes increase, but no change to simple underlying scaling laws with respect to requisite numbers of layers and layer thicknesses. These results establish a practical, generalizable framework for high-efficiency multimode coupling within the constraints of modern semiconductor foundries.
\end{abstract}

\maketitle

\section{Introduction}
Grating couplers offer potential interfaces for many free-space optical beams to scatter into many propagating modes for on-chip light manipulation. However, there is a dearth of designs that simultaneously support multiple wavelengths and incident angles, while satisfying foundry-imposed constraints on minimum feature size and fabrication tolerance. In this paper, we demonstrate examples of foundry-compatible grating couplers that can efficiently couple multiple independent waves from free space into waveguide modes on a chip, using a robust inverse design framework. We focus on grating couplers intended for visible (near 600 nm wavelengths) and near-infrared (near 900 nm wavelengths) wavelengths, where feature-size constraints are most stringent (fixed minimum sizes represent a larger fraction of the wavelength), thereby serving as a good testbed for robust inverse design and corresponding scaling laws. Incorporating minimum feature sizes typically increases the total computational time of the design process, but high-performance designs remain achievable. We demonstrate a number of examples in which four independent free-space waves, either at four frequencies for a single wavenumber or two frequencies for two wavenumbers, can be coupled via 2D grating couplers to the fundamental (and second, as needed) waveguide mode at average efficiencies in the range of 50--60\%, even, for example, with 62 nm minimum feature sizes and excitation wavelengths between 800--900 nm. Robust optimization further limits efficiency penalties to less than 5\% across all major types of fabrication imperfection: over/under-etch, critical-dimension variation, overlay misalignment, and sidewall angle deviations. Coupling efficiency decreases monotonically with minimum feature size, albeit relatively slowly, and we find an interesting solution to counteract this penalty: if one \emph{reduces} the index contrast, by filling the grating air holes with a material such as SiO$_2$, then higher average coupling efficiencies can be recovered even for moderate to large minimum feature sizes. 

Designing high-efficiency grating couplers is important for applications ranging from integrated photonics circuit~\cite{mekis2010grating,dai2012passive,marchetti2019coupling} and optical interconnects~\cite{shi2020scaling,yang2022multi,sun2015single} to quantum computing~\cite{mehta2016integrated,niffenegger2020integrated,mehta2020integrated,blumenthal2024enabling}. There is a relatively mature theoretical understanding of single-function grating couplers, including fundamental phase-matching conditions~\cite{tamir1977analysis,petit1980tutorial}, apodization of the fill factor or grating period for wavefront shaping~\cite{chen2010apodized,marchetti2017high,zhao2020design}, as well as more complex interactions, including guided-mode resonances~\cite{wang1990guided,rosenblatt1997resonant,quaranta2018recent} and periodic grating bands~\cite{tang2010highly}. These approaches have enabled a steady progression in single-function efficiency, from less than 40\% in early stage silicon-on-insulator (SOI) gratings~\cite{taillaert2002out,taillaert2006grating} to records approaching 90\%~\cite{ding2014fully,zaoui2014bridging}. Adding structural complexity through poly-silicon overlays~\cite{roelkens2006high,vermeulen2010high} or bi-layer structures~\cite{sacher2014wide} can increase directionality and bandwidth. Computational inverse design~\cite{su2018fully,sapra2019inverse,hammond2022multi} further extends the performance frontier beyond what analytical or parametric methods can reach. Near-unity coupling efficiencies can be achieved with inverse design~\cite{michaels2018inverse}, though potentially with increased fabrication challenges, such as many layers and/or small features. Wafer-scale experimental validations of inverse-designed grating couplers have been reported using 193~nm DUV immersion lithography~\cite{van2021wafer,van2022wafer}, demonstrating that adjoint-optimized designs are feasible in the context of industrial process variability, though these demonstrations targeted single-beam, single-wavelength operation.

On the other hand, multi-functional grating couplers, which simultaneously couple more than one beam, wavelength, or polarization, are significantly less explored. Early work by Ref.~\cite{taillaert2003compact} demonstrated a 2D grating that simultaneously couples and splits two orthogonal polarizations. Subsequently, conventional design methods extended these concepts to bi-wavelength polarization-splitting~\cite{streshinsky2013compact,cheng2022single}. Inverse design has pushed multi-functional performance further, producing grating couplers that couple two wavelengths~\cite{piggott2014inverse,su2018fully,sideris2022foundry}, as well as a four-wavelength coupler exploiting phase-matched resonant bands of a single Fourier component~\cite{su2024topology}. However, existing multi-functional designs have largely been limited to few functionalities, achieving moderate coupling efficiencies ($\lesssim 50\%$), with limited systematic analysis (e.g., scaling laws) of how foundry constraints affect multi-functional coupling performance. It remains an open question whether high coupling efficiencies can be achieved when scaling to four or more independent couplings with enforcement of minimum feature sizes and robustness to fabrication imperfections, particularly at visible and near-infrared wavelengths where minimum feature sizes on the order of 50--$\SI{100}{nm}$ can significantly affect performance.

\begin{figure*}[h]
    \begin{center}
    \includegraphics[width=1\linewidth]{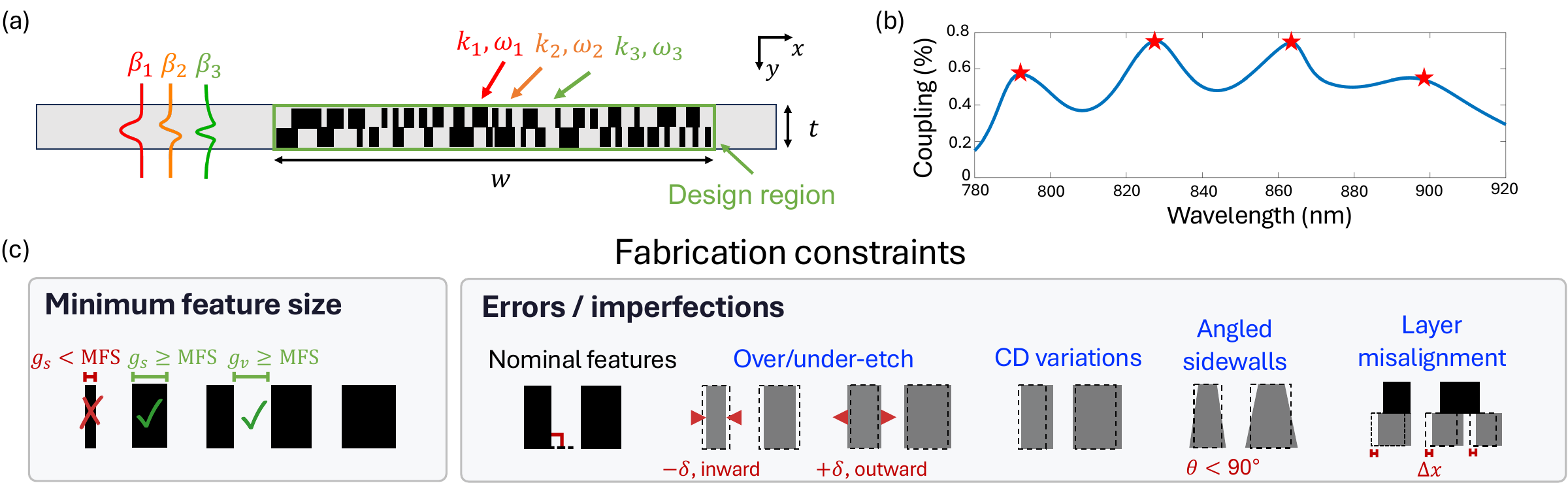}
    \end{center}
    \caption{Multi-functional grating coupler designs and considered fabrication imperfections. (a) A grating coupler simultaneously couples multiple incident free-space beams with distinct surface-parallel wavenumbers $k(\omega)$ into different waveguide modes with propagation constants $\beta(\omega)$. (Each wavenumber has a nonzero spread, ensuring finite-sized beams.) (b) Example spectrum of a 4-frequency, minimum-feature-size-compliant coupler, with distinct resonances at each target frequency. (c) Fabrication constraints considered in this work: minimum feature size (linewidth/linespacing) constraints, systematic over/under-etch ($\pm\delta$ on each side), random critical-dimension (CD) variations, angled sidewalls ($\theta < \ang{90}$), and layer misalignment ($\Delta x$).}
\label{fig:problem setup}
\end{figure*}

Throughout, we consider a bilayer grating coupler of a patterned design region with width $w$ and thickness $t$ (Fig.~\ref{fig:problem setup}(a)). It is designed to simultaneously couples $n_\lambda$ incident beams, each with center parallel wavenumber $k(\omega)$, into distinct waveguide modes with propagation constants $\beta(\omega)$. We consider minimum feature size constraints as well as a variety of fabrication imperfections: systematic over/under-etch, random critical dimension (CD) variations, angled sidewalls, and layer misalignment, as summarized in Fig.~\ref{fig:problem setup}(c). 

\section{Feature-size-compliant inverse design workflow}\label{sec:mfs_gc}
In this section, we describe our inverse design approach to designing multilayer grating couplers that satisfy minimum feature size constraints.  We find that minimum feature sizes on the order of 50--100 nm, for wavelengths in the visible and near infrared especially, are the strongest constraints on device performance, of the six  imperfections/tolerances depicted in Fig.~\ref{fig:problem setup}. The design proceeds in four stages: (1) gray-scale ``topology optimization'', (2) progressive binarization, (3) enforcement of minimum feature size constraints, and (4) edge-based shape optimization. This workflow enables only modest coupling efficiency reductions, although it seems to require some amount of testing over optimization hyperparameters. Tradeoff curves between best coupling efficiencies and minimum feature sizes show both features apparently independent of architectural details as well as features with interesting differences among different architectures. Perhaps the most interesting example of the latter is an improvement in average multimode coupling efficiency when replacing the air holes in our grating couplers with oxide materials. \emph{Reduced} refractive index contrast, typically a source of performance degradation, can reverse the penalties arising from minimum feature size constraints.

We start with gradient-based topology optimization~\cite{bendsoe1988generating,bendsoe2013topology,jensen2011topology}. Among the plethora of alternative design approaches now available (including machine learning~\cite{ma2021deep}, convex optimization~\cite{kuang2020computational,gertler2025many}, and global optimization~\cite{bennet2024illustrated}), it appears true that gradient-based design methods still offer the best combination of high device performance and (relatively) low computational cost. At its heart is the efficient computation of gradients of any objective with respect to all geometrical degrees of freedom with, essentially, two simulations (``forward'' and ``adjoint'') per independent excitation, independent of the number of design parameters. This powerful tool has enabled high-performance inverse-designed devices across a broad range of photonic platforms, including integrated photonics~\cite{yang2020inverse,yang2022multi}, metasurfaces~\cite{chung2023inverse,pestourie2018inverse}, photovoltaics~\cite{ganapati2013light}, imaging~\cite{lin2021end,li2022inverse}, and more. We apply this framework to design 2D multifunctional grating couplers. For a general grating coupler problem, one can simplify the designable volume to a rectangle or box, which can be further divided into ``layers'' within which the ``height'' degrees of freedom are fixed (e.g., vertical walls, slanted walls, etc.).

The requirement that fabrication produces ``binary'' materials (e.g., either SiN or air, with no in-between) can be addressed with three possible methods: ``level-set'' optimizations~\cite{wang2003level,van2013level}, which allow changes in shape or topology while retaining binary materials throughout the process, shape optimization~\cite{lalau2013adjoint}, in which only boundaries between binary materials are moved, preventing topology changes, or ``topology optimization''~\cite{jensen2005topology,Christiansen2021-qb}, in which the material parameters are relaxed to a continuum of grayscale variables, enabling both shape and topology changes across iterations, while requiring some type of binary enforcement by the end of the optimization procedure. Perhaps the strongest advantage of topology optimization over level-set optimization is the simplicity of implementation: level-set flow dynamics from a gradient are not trivial to implement, nor are their gradient computations with typical electromagnetic solvers. By contrast, grayscale materials enable simple gradient computations, and binarization can be achieved (at some computational cost) through appropriate penalization. We take this approach.

\begin{figure*}[tb]
    \begin{center}
    \includegraphics[width=1\linewidth]{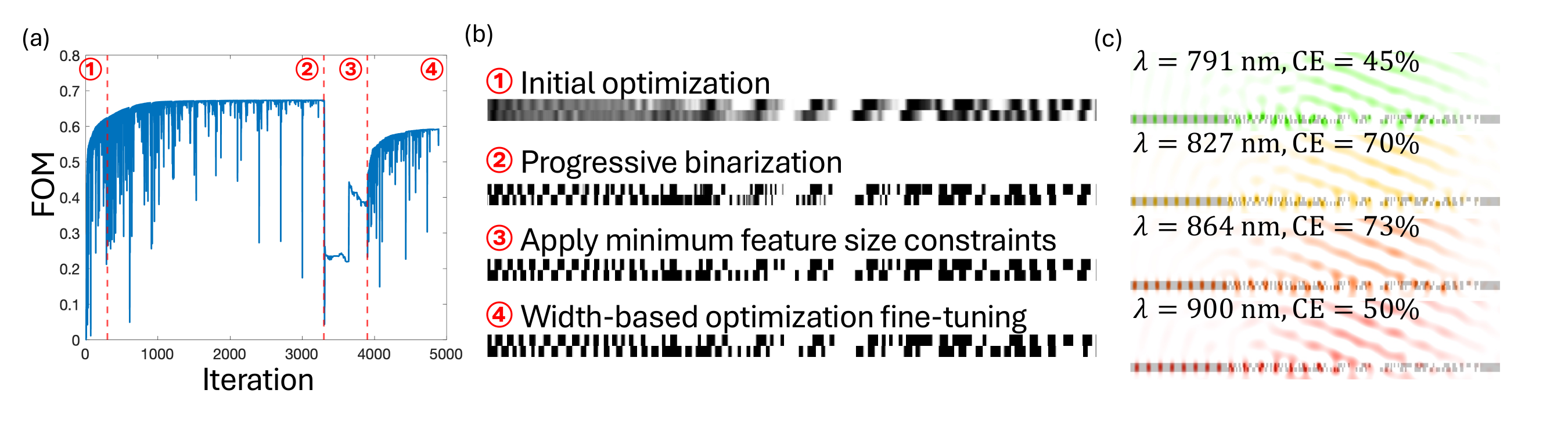}
    \end{center}
    \caption{Inverse-design workflow for a grating coupler with minimum feature size $\mfs = \SI{62}{nm}$, optimized over 4 wavelengths, $\lambda \in \{791, 827, 864, 900\}$~nm, at center parallel wavenumber $k \approx \SI{0.43}{\mu m^{-1}}$. (a) Optimization objective (FOM) versus iteration, with annotations indicating the transition between optimization stages. (b) Grating coupler geometries showing the four-stage optimization pipeline: (i) initial grayscale optimization, (ii) progressive binarization via smoothly increasing projection steepness $\beta$, (iii) applying $\SI{62}{nm}$ minimum-feature-size constraints, and (iv) geometric fine-tuning using an feature-width-based parameterization. (c) Resulting field patterns (real part of electric field, $\Re(\mathbf{E})$) for the optimized grating coupler in (b).}   
\label{fig:design_procedure}
\end{figure*}

The design process starts with gradient-based optimization of the grayscale density $\rho(\xv)$, from which the permittivity is written $\varepsilon(\xv) = \varepsilon_1 + (\varepsilon_2 - \varepsilon_1)\rho(\xv)$, with $\varepsilon_1$ the background material and $\varepsilon_2$ the higher-index material. Binarization of the permittivities requires forcing $\rho$ to 0 and 1 values only. The next step is to binarize the optimal grayscale design, for which several approaches have been developed, including penalty-based methods~\cite{jensen2005topology}, structural reparametrization~\cite{michaels2018inverse}, and projection-based binarization~\cite{wang2011projection,hammond2022high}. Here we adopt projection-based binarization as in Ref.~\cite{hammond2022high}, which proceeds in two steps. First, the raw density $\rho$ is smoothed by convolving with a low-pass filter, $\tilde{\rho} = w(\mathbf{r}) \circledast \rho$, to suppress high-spatial-frequency features below a scale set by the filter. We use a conic filter with radius $R$ for $w(\mathbf{r})$. The filtered field $\tilde{\rho}$ is then projected onto a smoothed Heaviside function, via the modified hyperbolic tangent function~\cite{lazarov2016length,chen2024validation,hammond2021photonic}:
\begin{equation}
    \bar{\rho} = \frac{\tanh(\beta\eta) + \tanh(\beta(\tilde{\rho} - \eta))}{\tanh(\beta\eta) + \tanh(\beta(1-\eta))},
    \label{eq:projection}
\end{equation}
where $\beta$ is a steepness parameter and $\eta$ is a threshold parameter. The relative permittivity is then interpreted with the modified permittivity, $\varepsilon_r(\bar{\rho}) = \varepsilon_1 + \bar{\rho}(\varepsilon_2 - \varepsilon_1)$. A binary structure is obtained by progressively increasing $\beta$ throughout the optimization. Standard ``soft binarization'' suffers from slow gradient convergence as $\beta$ increases and near-non-differentiability as $\beta\rightarrow\infty$. We adopt a recently developed subpixel-smoothed projection (SSP) method~\cite{hammond2025unifying}, which analytically accounts for partial filling of boundary pixels as $\beta \rightarrow \infty$, producing more accurate gradients at material interfaces and faster convergence. We find that choosing the right $\beta$ ``schedule'' is critical to minimizing performance loss during binarization. Ramping up $\beta$ overly aggressively traps the optimizer in a poor local optimum, whereas too gradual of an increase wastes computation time. Empirically, we adopt the schedule $\beta=[1,3,7,13,20,30,40,50,75,100,200, 500, \infty]$, with 400 gradient-descent steps per $\beta$, which we find sufficient to maintain high performance through the binarization stages. Fig.~\ref{fig:design_procedure}(a) shows an example of the evolution of an optimization figure of merit (``FOM''), average mode-coupling efficiency, as a function of iteration count for a grating-coupler optimization. The first two stages show the grayscale optimization process (``1'') followed by the binarization process (``2''). The former shows a rapid increase in FOM value, typically achieving high performance within a few hundred iterations, while the latter shows a slow increase and stabilization of the FOM. It is not guaranteed that the FOM does not decrease during this stage, and indeed we do run into such examples for some choices of hyperparameters.  

After binarization, the next step is to enforce minimum feature size constraints. Here we follow the method of Ref.~\cite{zhou2015minimum}. Minimum feature sizes refer to the smallest diameter (width, in 2D) of both the ``solid'' (high-permittivity) and ``void'' (low-permittivity) regions. For a grating these correspond to the minimum linewidth and minimum line spacing, respectively. We denote the minimum feature size ``$\mfs$,'' chosen to be equal for both solid and void regions. Ref.~\cite{zhou2015minimum} introduces two geometric constraints to enforce a minimum feature size. In the density filter of Eq.~\ref{eq:projection},  varying the projection threshold $\eta$ produces a continuous family of eroded and dilated realizations of a design. One can then prescribe a threshold range, $(\eta_d, \eta_e)$, such that no topology change across them implies a minimum feature size no smaller than $\mfs$. These conditions are formalized as two differentiable inequality constraints on the filtered density field $\tilde{\rho}$, for minimum linewidth $g^s\leq\epsilon$ and minimum line spacing $g^v\leq\epsilon$:
\begin{align}\label{eq:mfs}
    g^s &= \frac{1}{n}\sum_{i} I^{s}_i \!\left[\min\!\left\{\tilde{\rho}_i - \eta_e,\;0\right\}\right]^2 \leq \epsilon, \qquad g^v = \frac{1}{n}\sum_{i} I^{v}_i \!\left[\min\!\left\{\eta_d - \tilde{\rho}_i,\;0\right\}\right]^2 \leq \epsilon,
\end{align}
where $\epsilon$ is a small numerical relaxation value, and $I^s$ and $I^v$ are indicator functions that identify the inflection region of the solid and void phase, respectively: $I^{s} = \bar{\rho} \exp\!\left(-c|\nabla\tilde{\rho}|^2\right)$, and $I^{v} = (1-\bar{\rho})\exp\!\left(-c|\nabla\tilde{\rho}|^2\right)$. The decay-rate hyperparameter $c$ controls the spatial sharpness of the indicator functions. The thresholds $\eta_e$ and $\eta_d$ are analytically determined from the ratio $\mfs/R$ following Ref.~\cite{qian2013topological}. 

One might think these constraints should be imposed early in the design process, perhaps to ``coax'' the optimization towards feature-size-compliant designs. However, applying them too early in the binarization stage tends to trap the optimizer in poor quality local optima~\cite{zhou2015minimum}. Hence, we apply them later in the process, as the optimizer is approaching a binary design. (We also need not wait until perfect binarization.) We activate Eq.~\ref{eq:mfs} at a prescribed $\beta_\text{start}$, chosen from $\{75, 100, 200, 500, \infty\}$, and maintain it for all subsequent $\beta$ values. The remaining hyperparameters $\{\eta_e,c, a_k\}$ are inspired from Ref.~\cite{arrieta2025hyperparameter}, though we find that a modified range is more effective for our particular design problem and/or simulation tool:
\begin{equation}\label{eq:hyerpara_set}
    \eta_e \in \{0.75,\, 0.65,\, 0.60,\, 0.56\}, \quad 
    c \in \{64,\, 128\}R^2, \quad 
    \epsilon \in \{10^{-6},\, 10^{-7},\, 10^{-8}\}.
\end{equation}
For a given $\mfs$, we run a combinatorial sweep over $\{\beta_{\mathrm{start}}, \eta_e,c, a_k\}$ and select the best-performing combination for each design.

We typically find that enforcing the ``hard'' feature-size constraint can lead to a dramatic reduction in FOM, as depicted in the third stage of Fig.~\ref{fig:design_procedure}(a). The significant FOM drop, from ${\approx}60\%$ to ${\approx}20\%$, arises as sub-$\mfs$ features are eliminated or merged and the optimizer adjusts the remaining features to restore performance. However, the dramatic efficiency penalty is \emph{not} equivalent to, for example, moving much earlier in the process to a binary, feature-size-compliant design, perhaps with a similar starting FOM (say 20\%). Instead, these designs tend to be close in parameter space to high-performance, binary, feature-size-compliant designs. 

We find that a final shape optimization procedure, in which the design variables are the interface positions that can be moved according to their gradients, while preserving both binarization and feature-size constraints, enables a final improvement of the overall efficiency that tends to come relatively close to the efficiency of the best grayscale designs. For each layer, the binarized density profile is converted to a set of edge positions $\{p_1,p_2,\ldots,p_n\}$, that mark the boundaries between materials. A new density field is constructed from these edge positions with a sigmoid projection~\cite{wein2018combined,padhy2025photos},
\begin{equation}
    \rho_i = \frac{1}{1 + e^{-k_s \phi_i}},
    \label{eq:sigmoid}
\end{equation}
where $\phi_i$ is the signed distance from grid pixel $i$ to its nearest feature boundary (positive inside solid, negative inside void), and $k_s$ controls the sharpness of the material transition. This construction follows the geometry projection framework~\cite{norato2004geometry,norato2015geometry,wein2020review}, mapping explicit geometric parameters onto a fixed grid without remeshing, while retaining full differentiability with respect to the edge positions. (In the same spirit as the density relaxation for topology optimization.) We choose $k_s = 12.5/h$, approximately twice the value recommended in Ref.~\cite{wein2020review}, to achieve a sharper boundary representation that enables feature size control at sub-pixel resolution, as shown in Section~\ref{sec:robustness}. Adjoint gradients with respect to the density field are backpropagated through the sigmoid and signed-distance function chain to yield parameter gradients, which drive the optimizer (any first-order optimizer can be used; we mention our specific choice below). The minimum feature size is maintained by applying pairwise-edge constraints $p_{j+1} - p_j \geq \mfs$ throughout. By design, the device topology is fixed throughout this stage. 

To summarize, our optimization flow has four stages: (1) grayscale optimization, with  density values freely varying to discover optimal performance; (2) progressive binarization, with the parameter $\beta$ continually increased towards $\infty$, driving the design toward a fully binary permittivity distribution; (3) minimum feature size  constraints are imposed, per Eq.~\ref{eq:mfs}; (4) the binary structure is reparameterized at its boundaries (interfaces) and fine-tuned via shape optimization, to recover performance while maintaining the minimum feature size $\mfs$.

\section{Feature-size-compliant multimode grating couplers}
We apply this framework to 2D grating couplers, as schematically depicted in Fig.~\ref{fig:problem setup}(a). We take SiN as the core material ($\varepsilon_\mathrm{SiN} = 4.0$, a high index with transparency in the visible), with air or SiO$_2$ for the cladding. The rectangular design region has a width of $w = 10.7~\mu\text{m}$ and a thickness of $t \in [0.2, 1.0]~\mu\text{m}$. As a prototype for many-mode design problems, we focus on 4-function optimizations, targeting either four frequencies at a single (equal) wavenumber, or two frequencies, each at two wavenumbers. From a preliminary design exploration in Ref.~\cite{Pyvovar2026subm}, without robustness considerations, it is clear that designs with a single layer cannot (generically) achieve high efficiencies for four functions (with exceptions for special cases, such as a single phase-matching condition across all frequencies), whereas two or more layers can unlock high average efficiencies. We consider two-layer geometries, balancing performance and fabrication complexity. We consider operating wavelengths from visible to near-infrared range (500--900~nm), relevant for applications in optical inspection~\cite{zhu2022optical,orji2018metrology,den2016optical} and quantum computing and atom interrogation~\cite{mehta2016integrated,bruzewicz2019trapped}. Moreover, relative to telecommunications wavelengths (e.g., \SI{1550}{nm}), these wavelengths are more sensitive to feature-size constraints and fabrication imperfections (which are all larger fractions of the wavelength), serving as a good testbed for our numerical approach. We divide the 500--\SI{900}{nm} range into 12 equally spaced wavelengths and select design targets from this subset, focusing either near \SI{600}{nm} or near \SI{900}{nm}.

As shown in Fig.~\ref{fig:problem setup}(a), outside of the design region, the material is homogeneous, with the thickness $t$ now defining a waveguide supporting a discrete number of propagating modes. Each incident beam is a 2D TE-polarized Gaussian beam impinging at an angle determined by its central surface-parallel wavenumber, $k$. The beam is focused at the center of the designable region, with a waist $w_0 = \SI{3}{\mu m}$, corresponding to a diameter of $\SI{6}{\mu m}$, between 5--10 wavelengths in size. (This choice is primarily for computational speed to fully explore all architectures, thicknesses, grating variations, etc. Similar optimizations at larger beam sizes and design-region sizes produce similar results.) The device is optimized to simultaneously couple $n_\lambda$ wavelengths, each at $n_{k}$ incident angles, into distinct waveguide modes, for $N_{\rm fun} = n_\lambda n_{k}$ total ``functions.'' (In general, we expect any generic pairings of $k$ and $\lambda$ to exhibit similar scaling laws; there is nothing special about our ``outer product'' or ``grid'' choice.) We define our figure of merit (FOM) as the average coupling efficiency across all target couplings. There are a variety of equivalent expressions for mode-coupling efficiencies, which are all forms of overlap integrals between the total field ($\mathbf{E}^{i}$, for excitation index $i$) in the presence of the scattering coupler, and the target mode field ($\mathbf{H}_{\rm tar}^{j}$, for target waveguide propagating mode $j$). We use a numerically convenient version of this overlap~\cite{michaels2018leveraging},
\begin{equation}
    \mathrm{FOM} = \frac{1}{N_{\rm fun}}\sum_{ij}
    \left| \int \frac{1}{2} \mathbf{E}^i \times 
    (\mathbf{H}^j_\mathrm{tar})^* \, dS \right|^2.
    \label{eq:fom}
\end{equation}
In Eq.~\ref{eq:fom}, the indices $i$ and $j$ run over all target couplings, from incident field (indexed by $i$, power normalized to 1) to propagating mode (indexed by $j$, also power normalized to 1). The domain of the integral is a waveguide output plane. We simulate the scattering of each coupler by a volume integral equation method accelerated by a fast direct solver~\cite{xue2023fullwave}, which achieves near-linear scaling of computational time as a function of the long dimension of the waveguide. The simulation domain is discretized on a uniform grid with pixel size $h = 1/64~\mu\text{m} = 15.6~\text{nm}$, corresponding to $1/16^{\rm th}$ of the wavelength inside the highest-refractive-index medium at the shortest considered wavelength, and the design region shares the same discretization. The minimum feature sizes are chosen as integer multiples of the pixel size:
\begin{equation}\label{eq:all_mfs}
    \mfs \in \{31, 47, 62, 78, 94, 109, 125\}\ \mathrm{nm}.
\end{equation}
This range is compatible with \SI{193}{nm} DUV immersion lithography~\cite{mack2008fundamental}, which has produced sub-\SI{100}{nm} feature size in grating couplers on production wafers~\cite{van2021wafer,van2022wafer}. Of course, electron-beam lithography can resolve finer features~\cite{vieu2000electron}, albeit with less scalability.

Inverse design of highly efficient scattering structures can produce strong resonances, with high quality factors, that are less robust to fabrication imperfections. To suppress this effect, we introduce a small imaginary part to the frequency, $\Im \omega$, as a damping factor (equivalent to added material losses), which inhibits long optical paths and thus suppresses high-$Q$ resonances, making the designed structures more robust. We set $\Re(\omega)/\left(2\Im(\omega)\right)=1000$ throughout all four optimization stages. The designs are  further optimized for fabrication robustness as described in Section~\ref{sec:robustness}.
    
With the designable-region degrees of freedom defined (either in their initial grayscale form or their eventual edge-position representation), the objective function of Eq.~\ref{eq:fom}, gradients computed by the adjoint method~\cite{miller2012photonic}, and constraints such as those of Eq.~\ref{eq:mfs}, the optimization can proceed with any of a variety of first-order (gradient-only) optimization methods. We use the method of moving asymptotes (MMA)~\cite{svanberg1987method}, implemented via the \textsc{NLOpt} library~\cite{NLopt}, through its \textsc{Matlab} interface. A single forward simulation at one wavelength takes approximately 1.5 seconds on an 8-core Intel Xeon Platinum 8268 node. The full four-stage optimization for one design completes in approximately 20 hours. Combinatorial parameter sweeps over $\{\beta_\text{start}, \eta_e, c, a_k, \mfs\}$ are distributed as independent jobs on a computing cluster of such nodes, with each job handling one parameter combination. Fig.~\ref{fig:design_procedure}, which contains the best outcomes of 1,200,000 design runs, requires approximately 2 days on 120 nodes. 

In Fig.~\ref{fig:design_procedure}, we illustrate the full four-stage workflow on a representative design with $n_\lambda = 4$ wavelengths, $\lambda \in \{791, 827, 864, 900\}$~nm, at a single center parallel wavenumber $k = 1/2.3\approx \SI{0.43}{\mu m^{-1}}$, with minimum feature size $\mfs = 62$~nm and waveguide thickness $t=0.375$~um. For the minimum feature size constraints, we set $\beta_\text{start} = 500$, $\eta_e = 0.75$, $c = 64$, $a_k = 10^{-7}$, selected from the combinatorial sweep described above. The bilayer SiN/air grating coupler is optimized to couple all four beams into the fundamental waveguide mode. Figure~\ref{fig:design_procedure}(a) shows the average coupling efficiency FOM as a function of iteration, and Fig.~\ref{fig:design_procedure}(b) shows the corresponding grating structure at the end of each stage. In stage~(i), optimization rapidly increases the FOM to approximately 63\%, albeit in physically unrealistic grayscale structures (cf. top row of Fig.~\ref{fig:design_procedure}(b)). Stage~(ii) progressively increases binarization. Each jump in the value of $\beta$ causes the FOM to drop, followed by a recovery. Due to the effective $\beta$ schedule (found by trial and error), the performance ultimately retains an FOM close to the grayscale optimum. The resulting structure (second row of Fig.~\ref{fig:design_procedure}(b)) is nearly binary but still contains narrow features that violate the \SI{62}{nm} minimum feature size. In stage~(iii), activating the geometric constraints of Eq.~\ref{eq:mfs} causes an immediate, hard-to-avoid drop in the FOM; in this case, of approximately 40\%, as too-small features are automatically merged or eliminated. In stage~(iv), the edge-based width fine-tuning recovers a substantial fraction of this loss by adjusting feature boundaries, at sub-pixel resolution, while preserving the device topology and feature-size compliance. The final structure (bottom row of Fig.~\ref{fig:design_procedure}(b)) differs from stage~(iii) only through a modest number of edge shifts, but boosts the FOM back to 59.5\%, retaining 90\% of the unconstrained grayscale performance while achieving binarization and feature-size compliance. The resulting grating exhibits a complex, aperiodic pattern qualitatively distinct from classic periodic or apodized designs, reflecting the multi-functional coupling requirement.

Fig.~\ref{fig:design_procedure}(c) shows the per-wavelength field patterns, $\mathrm{Re}(\mathbf{E})$, and coupling efficiencies for the final design. The grating efficiently couples the four wavelengths at efficiencies of 45\%, 70\%, 73\%, and 50\% at $\lambda = 791$, 827, 864, and 900~nm, respectively. The color-coded field patterns show each beam coupling into the fundamental waveguide mode; the distinct scattering patterns within the design region reflect the wavelength-dependent interference of each Gaussian beam by the structure. 

\begin{figure*}[tb]
\begin{center}
\includegraphics[width=1\linewidth]{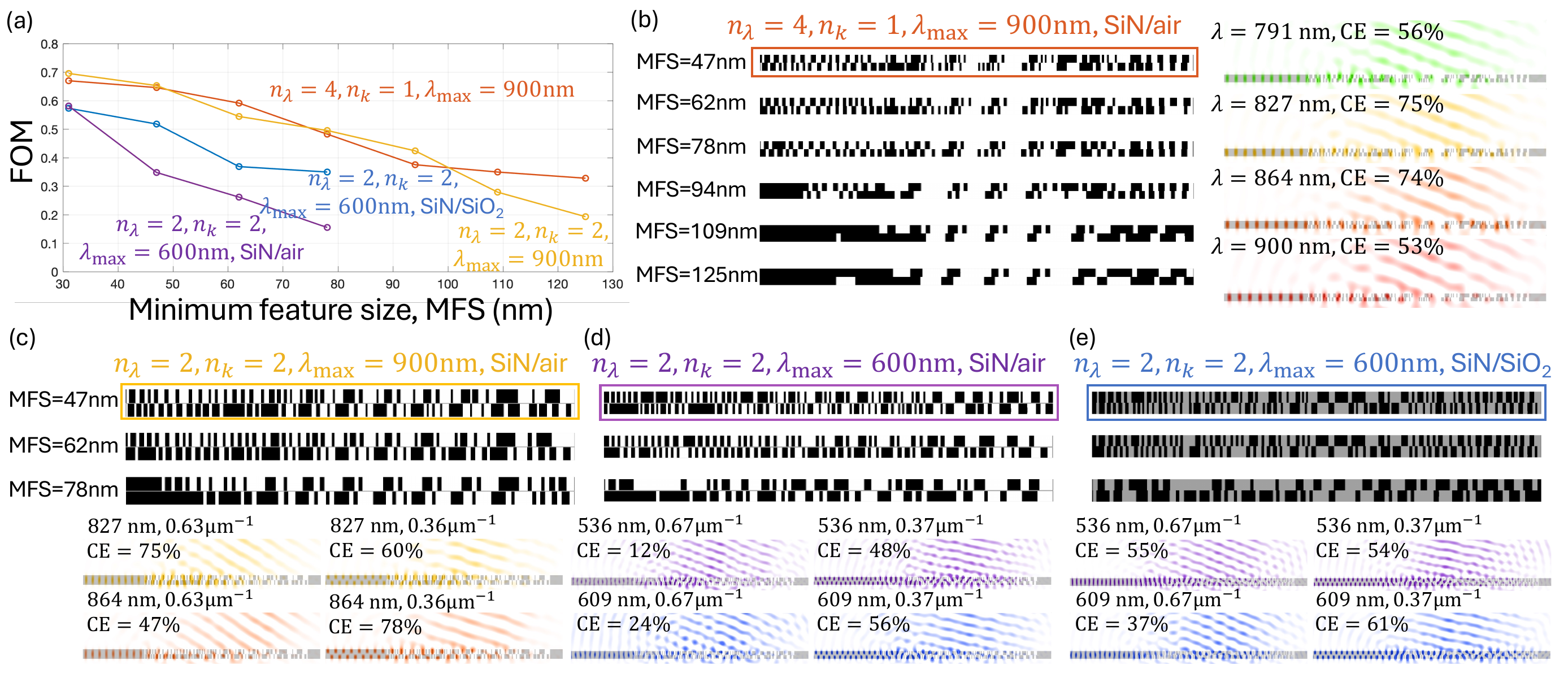}
\end{center}
\caption{Performance tradeoff with minimum feature size for various design configurations. (a) Best achievable FOM (average coupling efficiency) versus imposed minimum feature size $\mfs$ for multiple design configurations, including the number of coupled beams $n_\lambda$, the number of parallel wavenumber $n_{k}$, the maximum operating wavelength $\lambda_{\mathrm{max}}$ and the material platforms. (b–e) Representative optimized binary grating couplers at various minimum feature sizes ($\mfs = 47, 62, 94, 109, 125~\mathrm{nm}$) in (a). For each configuration, the corresponding field patterns, $\Re(\mathbf{E})$, are shown for the designs with $\mfs=47~\mathrm{nm}$.}
\label{fig:mfs scaling law}
\end{figure*}
We perform similar minimum-feature-size numerical experiments for three additional configurations beyond the $n_\lambda = 4$, $n_{k} = 1$, $\lambda_\mathrm{max} = 900$~nm baseline: (1) $n_\lambda = 2$, $n_{k} = 2$ ($k=1/1.6\approx \SI{0.63}{\um^{-1}}, 1/2.8\approx \SI{0.36}{\um^{-1}}$) at $\lambda = 827, 864$~nm, design region thickness $t=0.625~\mu\mathrm{m}$, with a bilayer SiN/air design and a 16~nm SiO$_2$ spacer; (2) $n_\lambda = 2$, $n_{k} = 2$ ($k=1/1.5\approx 0.67, 1/2.7\approx 0.37~\mu\mathrm{m}^{-1}$) at $\lambda = 536, 609$~nm, design region thickness $t=0.5~\mu\mathrm{m}$, same bilayer SiN/air configuration; (3) identical to (2) but with SiN/SiO$_2$ in place of SiN/air and no spacer layer. For each configuration, we sweep the $\mfs$ over the values in \eqref{all_mfs}, run the full hyperparameter combinatorial sweep of \eqref{hyerpara_set}, and select the best achievable FOM at each $\mfs$. The four resulting tradeoff curves are shown in Fig.~\ref{fig:mfs scaling law}(a), with representative structures and field patterns at $\mfs = 47$~nm in Fig.~\ref{fig:mfs scaling law}(b--e).

It is worth noting that the designs exhibit peaked spectral response concentrated near the target wavelengths. The $n_\lambda=4$, $n_{k}=1$, $\lambda_\mathrm{max}=900$~nm design at $\mfs=47$~nm achieves a mean coupling efficiency of ${\approx}65\%$, and the wavelength-resolved spectrum shows four distinct peaks centered near the target wavelengths, as in Fig.~\ref{fig:problem setup}(b). In contrast to broadening a single phase-matching condition~\cite{sapra2019inverse}, our designs must effectively support effective phase-matching and apodization for four separate excitations. The frequency bandwidths of each efficiency peak can be numerically controlled by altering the imaginary part of the frequency, discussed above, and by changing the diameter of the grating coupler.

As expected, all tradeoff curves in Fig.~\ref{fig:mfs scaling law}(a) decrease monotonically as $\mfs$ increases, although the rate of decay varies with operating wavelength and material system. Physically, as $\mfs/\lambda$ increases, the device progressively loses sufficient geometric ``tuning knobs'' and control. This trend is consistent with the reported results in Ref.~\cite{michaels2018inverse}, though the efficiency decays more rapidly with $\mfs/\lambda$ in our designs, which is likely due to the 4X increase in the number of targeted couplings.

The $n_\lambda = 4$, $n_{k} = 1$ and $n_\lambda = 2$, $n_{k} = 2$ configurations at $\lambda_\mathrm{max} = 900$~nm produce nearly identical FOM tradeoff curves against $\mfs$, as in Fig.~\ref{fig:mfs scaling law}(a). This suggests that the key determinant of the FOM is the total targeted number of couplings (mode pairs), independent of whether they are spatial or spectral channels. This corroborates theoretical predictions of Ref.~\cite{miller2013complicated}, showing that the minimum volume of a linear optical device scales with the total number of input-output mode pairs, regardless of whether those modes are distinguished spectrally or spatially. This suggests that a four-wavelength coupler and a two-wavelength, two-angle coupler have the same functional design complexities, when operating within similar wavelength ranges. Another interesting finding is that the addition of a $16$~nm SiO$_2$ spacer layer ($\varepsilon_\mathrm{SiO2}=2.1$) does not measurably affect the FOM tradeoff with $\mfs$, confirming that performance is primarily determined by the total number of functions.

Reducing the operating wavelength from $\lambda_\mathrm{max} = \SI{900}{nm}$ to \SI{600}{nm} at fixed $n_\lambda \times n_{k} = 4$, for fixed materials, shifts the entire scaling curve downward as in Fig.~\ref{fig:mfs scaling law}(a). At the tail ends of the curves, the difference is approximately a factor of 2/3 in $\mfs$: the \SI{900}{nm}-wavelength-range configuration at a given $\mfs$ (e.g., \SI{120}{nm}) achieves a similar FOM as the \SI{600}{nm}-wavelength-range configuration (e.g., near $\mfs \approx \SI{80}{nm}$), approximately matching the wavelength ratios. 

Of particular interest in Fig.~\ref{fig:mfs scaling law}(a), we find that introducing SiO$_2$ as the filler material in the SiN grating (for $\lambda_\mathrm{max} = \SI{600}{nm}$) outperforms the SiN/air configuration across most of the $\mfs$ range, despite its lower refractive index contrast ($\Delta n \approx 0.55$ for SiN/SiO$_2$ versus $\approx 1.0$ for SiN/air). A natural concern with the SiN/SiO$_2$ platform is the lower index contrast, limiting the scattering strength of each grating pitch, making it difficult to match the incident Gaussian beam profile and thereby potentially reducing coupling efficiency~\cite{korvcek2023low}. This associated problem is typically remedied by enlarging the device footprint~\cite{sacher2014wide,ong2018sinx,ruiz2025inverse} or employing back reflectors~\cite{romero2013visible,hong2019high,nambiar2019high}. The unconstrained grayscale design confirm this expectation: SiN/air achieves ${\approx}66\%$ mean coupling efficiency at the end of stage~(1), compared to ${\approx}60\%$ for SiN/SiO2 (not shown in the figure). However, once minimum feature size constraints are applied, this ordering reverses: the SiN/SiO2 design begins to outperform SiN/air for $\mfs > \SI{31}{nm}$, for the same design footprint. The distinction is even more dramatic at $\mfs=\SI{47}{nm}$, as in Fig.~\ref{fig:mfs scaling law}(d,e): the SiN/air design achieves mean coupling efficiencies of 35\% across four functionalities, whereas the SiN/SiO2 design achieves 52\%, a 1.5X improvement. Larger feature sizes as a fabrication constraint force uncontrollably strong scattering, without sufficient geometric degrees of freedom to tame it. Reduced contrast offers a means to weaken this scattering, offering more granular control over the scattering wavefronts. Of course, one would not want to push this too far, or the scattering strengths would weaken too much to be recoverable.

In addition, reducing the index contrast improves the effective fabrication tolerance~\cite{pita2025integrated,baets2016silicon}. A dimensional error $\delta w$ in a grating tooth produces a phase error that scales with $\Delta n$: the same width deviation causes a proportionally larger effective-index shift in SiN/air than in SiN/SiO2. As the minimum feature size increases, per-feature phase errors accumulate more severely across the grating in SiN/air design, greatly degrading performance. Taken together, these two mechanisms suggest that for designs where $\mfs/\lambda$ is inherently large, e.g., at visible and near-infrared wavelengths, filling air voids with SiO$_2$ (or another material) is a practical route to recovering coupling efficiency without requiring finer features.

\section{Robust, foundry-compatible optimal designs}\label{sec:robustness}
Building on the feature-size enforcement of the previous section, in this section we describe how to ensure that designs retain high performance when subjected to realistic manufacturing variations/imperfections. Semiconductor fabrication processes introduce a variety of unavoidable variations, typically normalized to the minimum feature size (referred to as ``critical dimension,'' or ``CD,'' in this section)~\cite{mack2008fundamental,den2016optical}. As depicted in Fig.~\ref{fig:problem setup}(c), we consider four imperfections: (1) a uniform over- or under-etch of all feature widths by $\pm\delta$, within the range of $\pm10\%$ of CD; (2) random width perturbations, for each feature, with amplitudes of up to $10\%$ of CD; (3) sidewall angles $\theta$ that deviate from perfectly vertical, $\ang{80}\leq\theta \leq \ang{90}$; (4) relative overlay misalignment $\Delta x$, within $\pm10\%$ of CD. We find the strongest sensitivity to systematic etch errors, whereas our designs are inherently robust over all other 3 types of imperfections. 
 
The edge-based shape parameterization of Section~\ref{sec:mfs_gc} makes robustness testing and optimization straightforward. At the end of stage (4), the optimized multi-layer grating coupler is represented by a set of edge positions $\{p_1,p_2,\ldots,p_n\}$ denoting the material boundaries. Since they are continuous-valued real numbers, feature-width perturbations can be introduced at length scales well below the pixel size $h$, without any re-meshing or re-parameterization required. 

\begin{figure*}[tb]
    \begin{center}
    \includegraphics[width=1\linewidth]{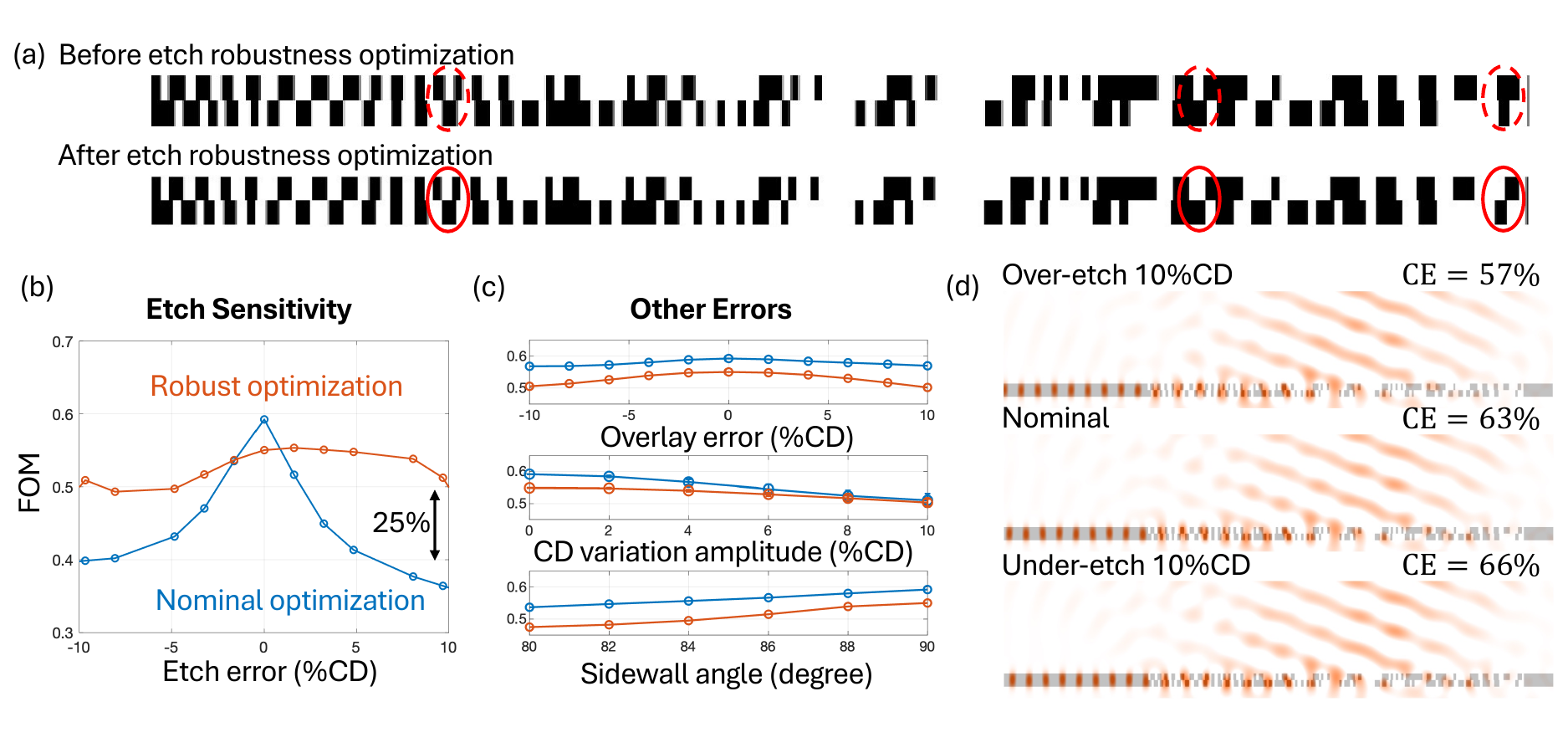}
    \end{center}
    \caption{Robustness testing and optimization of the bilayer grating coupler of Fig. \ref{fig:design_procedure}. (a) Optimized grating coupler structures before (top) and after (bottom) over/under-etch robustness optimization. (b) FOM versus systematic etch error (\%CD) for nominally optimized (blue, top in (a)) and robustness-optimized (orange, bottom in (a)) designs; the robustness-optimized design recovers ${\approx{25}}\%$ of the etch-induced efficiency loss at large etch errors. (c) FOM sensitivity of other fabrication error types: overlay misalignment error, random CD variation, and sidewall angle, for both nominal (top in (a)) and etch robustness-optimized (bottom in (a)) designs, showing the robustness-optimized design retains good performance across all major fabrication variations. (d) Field patterns ($\Re(\mathbf{E})$) of the etch-robustness-optimized design at $\lambda = \SI{864}{nm}$ under three etch conditions: over-etch ($-10\%$~CD, CE~$= 57\%$), nominal (CE~$= 63\%$), and under-etch ($+10\%$~CD, CE~$= 66\%$).}
\label{fig:robustness opt}
\end{figure*}
We first characterize the sensitivity of the nominally optimized (non-robust) device designed for $n_\lambda = 4$, $n_{k} = 1$, $\lambda_\mathrm{max} = 900$~nm, and $\mfs = \SI{62}{nm}$, from Fig.~\ref{fig:design_procedure}. As shown in Fig.~\ref{fig:robustness opt}(b), this device is highly sensitive to systematic etch errors: the peak efficiency of 59.5\% at nominal conditions drops to near or below 40\% at $\pm10\%$\,CD etch error. Such extreme sensitivity to small etch error makes it a great challenging for reliable fabrication: $\pm2\%$CD error is required for this device to support performance greater than $50\%$.

In contrast, this device exhibits substantially less sensitivity to the other three error types, as in Fig.~\ref{fig:robustness opt}(c): performance degrades by only 5\%, 10\%, and 6\% over the considered ranges of overlay error, random CD variation, and sidewall angle, respectively. The difference between systematic and random errors is physically intuitive: a systematic over- or under-etch produces a uniform shift of all feature edges, which results in a larger total change in material volume and hence a stronger collective perturbation. Random CD variations, by contrast, introduce independent positive, zero, or negative edge displacements across the device, leading to smaller total volume changes for the same nominal range, reducing the overall perturbation strength.

To mitigate the dominant etch sensitivity, we modify our edge-based width optimization at stage (4) to account for systematic geometric errors. We apply robustness optimization after stage~(4), rather than at earlier stages, waiting until the device topology is fixed. The edge parameterization naturally accommodates uniform width perturbations as simple shifts of each $p_j$, making the etch perturbation easy to evaluate without re-meshing. Instead of computing the FOM and its gradient only for the nominal geometry, we also evaluate two additional configurations: over-etch and under-etch of every feature by $\delta$ at each iteration, and then optimize their weighted mean:
\begin{equation}
    \overline{\mathrm{FOM}} = \sum_{q} w_q\mathrm{FOM}(\varepsilon_q),\quad \nabla\overline{\mathrm{FOM}}=\sum_{q} w_q\nabla\overline{\mathrm{FOM}}(\varepsilon_q),
\end{equation}
where $\varepsilon_q$ represents the material distribution for the $q$-th case (nominal, over-etch, under-etch). The weights $w_q$ allow tuning the relevant contributions of each configuration in the robustness enforcement; we choose $w_{\mathrm{nominal}}=1/2,w_{\mathrm{under}}=1/4,w_{\mathrm{over}}=1/4$, as in Ref.~\cite{wang2019robust}, with a target robustness range of $\pm10\%$ of CD.

As shown in Fig. \ref{fig:robustness opt}(a), the robustness-optimized design has only modest geometric adjustments relative to the nominal design, yet these modifications produce a remarkably robust performance across systematic etch errors, shown in Fig. \ref{fig:robustness opt}(b). The robust design maintains an average coupling efficiency near or above 50\% across all four target wavelengths over the full range of $\pm10\%$ of CD. The peak performance at nominal condition drops by 5\% from non-robust design to 55\%, a modest cost in return for robustness. The robust design achieves $\approx25\%$ more performance at $10\%$ etch error relative to the nominal design. Devices designed for larger range of robustness is also possible by tuning the parameter $\delta$. 

Although the robustness optimization targets only systematic etch variations, the resulting design also exhibits reduced sensitivity to random CD errors. As shown in Fig.~\ref{fig:robustness opt}(c), the etch-robust design drops only 5\% efficiency for random variations of up to $10\%$ of CD, compared to 10\% for the nominal design. Devices designed to be robust for systematic edge deviations appear to also be robust to random deformations~\cite{wang2019robust}, implying that optimizing for systematic etch errors forces the design away from sharp performance peaks in the geometric parameter space and onto a flatter region of the landscape that is inherently less sensitive to geometric perturbations of any kind. For overlay misalignment and sidewall angle, both the nominal and robust designs exhibit comparable sensitivity (5\% and 6\%, respectively), indicating that the bilayer grating geometry is inherently tolerant to these error types. Overlay shifts displace the two layers laterally with respect to each other, introducing a distributed perturbation to the scattering profile. We speculate that the relatively weak observed sensitivity may arise from partial cancelations of the induced phase and amplitude perturbations across the device. Sidewall angle deviations modify the grating ``tooth'' shape but retain the effective fill fraction of scattering element, thus also only weakly affecting coupling efficiencies in our designs.

\section{Discussion and Outlook}
We have demonstrated a foundry-compatible inverse-design framework for multifunctional grating couplers. Our approach involves four stages: grayscale topology optimization, progressive binarization, minimum-feature-size enforcement, and edge-based shape refinement. This process, with tuned hyperparameters (and some amount of repeated optimization with different initial designs), consistently recovers most of the performance of unconstrained grayscale designs while satisfying binary-fabrication requirements. Leveraging this workflow, we demonstrated bilayer SiN/air grating couplers that simultaneously couple four independent incident channels into guided modes at mean efficiencies in the 50--60\% range, for minimum feature sizes of \SI{62}{nm} at wavelengths near 900~nm. Robustness optimization substantially suppresses sensitivity to systematic over/under-etch error while incurring only modest efficiency penalties, and maintains robustness to critical-dimension variation, overlay misalignment, and sidewall-angle deviations. Across all configurations studied, the total number of input--output functionalities and the operating wavelength emerges are the primary determinants of the minimum-feature-size tradeoff. We also identified a counterintuitive but practically useful result: \emph{reducing} the refractive-index contrast in the design region can \emph{improve} performance at moderate to large minimum feature sizes. 

Looking forward, there are a number of avenues for further exploration. The current workflow relies on a combinatorial sweep over binarization and constraint hyperparameters ($\beta_\mathrm{start}$, $\eta_e$, $c$, $a_k$), which is time-consuming. One can envision reducing this reliance through a systematic study of optimal hyperparameter choices as a function of design configurations, including the number of layers, operating wavelength, and minimum feature size. (Ref.~\cite{arrieta2025hyperparameter} offers one such study, but for our scenarios and solvers we did not find their settings optimal, highlighting the difficulty of achieving a general, hyperparameter-free approach.) An alternative avenue is to bypass the density-based stages entirely, and applying edge-based shape optimization from the outset. This approach faces two challenges: the local optimum appears to depend strongly on the initial structure, and it cannot easily modify the topology (i.e., inserting or removing features). The first challenge can be addressed with physics-informed initial geometries, for example apodized gratings with local periodicity satisfying the phase-matching condition. The second can be alleviated with topological insertions and deletions through the design process~\cite{eschenauer1994bubble,amstutz2006new,allaire2005structural,miller2012photonic}. If both challenges can be overcome, the computational cost of producing a fabrication-ready design may be reduced by up to an order of magnitude.

The computational cost of the adjoint-based design process is directly proportional to the computational time of each simulation. In this work, it takes $\approx2$ seconds per wavelength on a single CPU core using our VIE solver~\cite{xue2023fullwave}. For the 2D grating geometries considered here, this cost is manageable, whereas
scaling to larger design regions or higher-dimensional problems will require faster solvers. Modern acceleration strategies include GPU-accelerated FDTD solvers~\cite{hughes2021perspective} and neural-network surrogate models~\cite{ma2021deep}. In 3D, computational demands grow substantially. GPU-accelerated FDTD solvers such as Tidy3D~\cite{tidy3d} offer one path forward, while fast integral equation methods with improved compression or GPU-accelerated solvers may be essential to keep the full optimization pipeline tractable. 

Finally, the designs presented here target four input--output mode pairs in 2D, but preliminary 3D results suggest that coupling hundreds to thousands of modes may be possible~\cite{Pyvovar2026subm}. It remains an open question whether fabrication-friendly designs with comparable per-mode efficiencies can be discovered at these scales. The $\mfs$ tradeoff curves in Fig.~\ref{fig:mfs scaling law} provide a baseline. We expect the same feature-size findings to translate to many-many coupling in 3D, but it is important to computationally and experimentally validate the ultimate limits of high-efficiency, many-mode couplers.

\begin{acknowledgments}
This work was partially supported by grants from AFOSR (grant no. FA9550-24-1-0193) and the Simons Collaboration on Extreme Wave Phenomena Based on Symmetries (award no. SFI-MPS-EWP-00008530-09).
\end{acknowledgments}

\bibliography{main}

@ARTICLE{Christiansen2021-qb,
  title     = "Inverse design in photonics by topology optimization: tutorial",
  author    = "Christiansen, R E and Sigmund, O",
  journal   = "Journal of the Optical Society of America B",
  publisher = "opg.optica.org",
  volume    =  38,
  number    =  2,
  pages     = "496--509",
  year      =  2021
}

@article{Pyvovar2026subm,
  title={Many-mode grating couplers by avoiding undesired couplings},
  author={Pyvovar, Nazar and Li, Hao and Dai, Zhaowei and Miller, Owen D.},
  journal={In preparation},
  year={2026}
}

@article{hammond2025unifying,
  title={Unifying and accelerating level-set and density-based topology optimization by subpixel-smoothed projection},
  author={Hammond, Alec M and Oskooi, Ardavan and Hammond, Ian M and Chen, Mo and Ralph, Stephen E and Johnson, Steven G},
  journal={Optics Express},
  volume={33},
  number={16},
  pages={33620--33642},
  year={2025},
  publisher={Optica Publishing Group}
}

@article{hammond2021photonic,
  title={Photonic topology optimization with semiconductor-foundry design-rule constraints},
  author={Hammond, Alec M and Oskooi, Ardavan and Johnson, Steven G and Ralph, Stephen E},
  journal={Optics Express},
  volume={29},
  number={15},
  pages={23916--23938},
  year={2021},
  publisher={Optical Society of America}
}

@article{chen2024validation,
  title={Validation and characterization of algorithms and software for photonics inverse design},
  author={Chen, Mo and Christiansen, Rasmus E and Fan, Jonathan A and I{\c{s}}iklar, G{\"o}ktu{\u{g}} and Jiang, Jiaqi and Johnson, Steven G and Ma, Wenchao and Miller, Owen D and Oskooi, Ardavan and Schubert, Martin F and others},
  journal={Journal of the Optical Society of America B},
  volume={41},
  number={2},
  pages={A161--A176},
  year={2024},
  publisher={Optica Publishing Group}
}

@article{arrieta2025hyperparameter,
  title={Hyperparameter-free minimum-lengthscale constraints for topology optimization},
  author={Arrieta, Rodrigo and Romano, Giuseppe and Johnson, Steven G},
  journal={arXiv preprint arXiv:2507.16108},
  year={2025}
}

@article{xue2023fullwave,
  title={Fullwave design of cm-scale cylindrical metasurfaces via fast direct solvers},
  author={Xue, Wenjin and Zhang, Hanwen and Gopal, Abinand and Rokhlin, Vladimir and Miller, Owen D},
  journal={arXiv preprint arXiv:2308.08569},
  year={2023}
}

@article{hammond2022high,
  title={High-performance hybrid time/frequency-domain topology optimization for large-scale photonics inverse design},
  author={Hammond, Alec M and Oskooi, Ardavan and Chen, Mo and Lin, Zin and Johnson, Steven G and Ralph, Stephen E},
  journal={Optics Express},
  volume={30},
  number={3},
  pages={4467--4491},
  year={2022},
  publisher={Optical Society of America}
}

@phdthesis{miller2012photonic,
  title={Photonic design: From fundamental solar cell physics to computational inverse design},
  author={Miller, Owen Dennis},
  year={2012},
  school={University of California, Berkeley}
}

@article{jensen2011topology,
  title={Topology optimization for nano-photonics},
  author={Jensen, Jakob S{\o}ndergaard and Sigmund, Ole},
  journal={Laser \& Photonics Reviews},
  volume={5},
  number={2},
  pages={308--321},
  year={2011},
  publisher={Wiley Online Library}
}

@article{chung2023inverse,
  title={Inverse design of high-NA metalens for maskless lithography},
  author={Chung, Haejun and Zhang, Feng and Li, Hao and Miller, Owen D and Smith, Henry I},
  journal={Nanophotonics},
  volume={12},
  number={13},
  pages={2371--2381},
  year={2023},
  publisher={De Gruyter}
}

@article{pestourie2018inverse,
  title={Inverse design of large-area metasurfaces},
  author={Pestourie, Rapha{\"e}l and P{\'e}rez-Arancibia, Carlos and Lin, Zin and Shin, Wonseok and Capasso, Federico and Johnson, Steven G},
  journal={Optics express},
  volume={26},
  number={26},
  pages={33732--33747},
  year={2018},
  publisher={Optical Society of America}
}

@article{michaels2018inverse,
  title={Inverse design of near unity efficiency perfectly vertical grating couplers},
  author={Michaels, Andrew and Yablonovitch, Eli},
  journal={Optics express},
  volume={26},
  number={4},
  pages={4766--4779},
  year={2018},
  publisher={Optical Society of America}
}

@article{wang2011projection,
  title={On projection methods, convergence and robust formulations in topology optimization},
  author={Wang, Fengwen and Lazarov, Boyan Stefanov and Sigmund, Ole},
  journal={Structural and multidisciplinary optimization},
  volume={43},
  number={6},
  pages={767--784},
  year={2011},
  publisher={Springer}
}

@article{lazarov2016length,
  title={Length scale and manufacturability in density-based topology optimization},
  author={Lazarov, Boyan S and Wang, Fengwen and Sigmund, Ole},
  journal={Archive of Applied Mechanics},
  volume={86},
  number={1},
  pages={189--218},
  year={2016},
  publisher={Springer}
}

@article{qian2013topological,
  title={Topological design of electromechanical actuators with robustness toward over-and under-etching},
  author={Qian, Xiaoping and Sigmund, Ole},
  journal={Computer Methods in Applied Mechanics and Engineering},
  volume={253},
  pages={237--251},
  year={2013},
  publisher={Elsevier}
}

@article{zhou2015minimum,
  title={Minimum length scale in topology optimization by geometric constraints},
  author={Zhou, Mingdong and Lazarov, Boyan S and Wang, Fengwen and Sigmund, Ole},
  journal={Computer Methods in Applied Mechanics and Engineering},
  volume={293},
  pages={266--282},
  year={2015},
  publisher={Elsevier}
}

@article{norato2004geometry,
  title={A geometry projection method for shape optimization},
  author={Norato, J and Haber, R and Tortorelli, D and Bends{\o}e, Martin P},
  journal={International Journal for Numerical Methods in Engineering},
  volume={60},
  number={14},
  pages={2289--2312},
  year={2004},
  publisher={Wiley Online Library}
}

@article{norato2015geometry,
  title={A geometry projection method for continuum-based topology optimization with discrete elements},
  author={Norato, Julian A and Bell, B Kꎬ and Tortorelli, Daniel A},
  journal={Computer Methods in Applied Mechanics and Engineering},
  volume={293},
  pages={306--327},
  year={2015},
  publisher={Elsevier}
}

@article{wein2020review,
  title={A review on feature-mapping methods for structural optimization},
  author={Wein, Fabian and Dunning, Peter D and Norato, Juli{\'a}n A},
  journal={Structural and multidisciplinary optimization},
  volume={62},
  number={4},
  pages={1597--1638},
  year={2020},
  publisher={Springer}
}

@article{wein2018combined,
  title={A combined parametric shape optimization and ersatz material approach},
  author={Wein, Fabian and Stingl, Michael},
  journal={Structural and Multidisciplinary Optimization},
  volume={57},
  number={3},
  pages={1297--1315},
  year={2018},
  publisher={Springer}
}

@article{padhy2025photos,
  title={Photos: topology optimization of photonic components using a shape library},
  author={Padhy, Rahul Kumar and Chandrasekhar, Aaditya},
  journal={Engineering with Computers},
  volume={41},
  number={2},
  pages={1141--1153},
  year={2025},
  publisher={Springer}
}

@article{svanberg1987method,
  title={The method of moving asymptotes—a new method for structural optimization},
  author={Svanberg, Krister},
  journal={International journal for numerical methods in engineering},
  volume={24},
  number={2},
  pages={359--373},
  year={1987},
  publisher={Wiley Online Library}
}

@misc{NLopt,
  title = {The {NLopt} nonlinear-optimization package},
  author = {Steven G. Johnson},
  year = {2007},
  howpublished = {\url{https://github.com/stevengj/nlopt}}
}

@article{miller2013complicated,
  title={How complicated must an optical component be?},
  author={Miller, David AB},
  journal={Journal of the Optical Society of America A},
  volume={30},
  number={2},
  pages={238--251},
  year={2013},
  publisher={Optical Society of America}
}

@article{sacher2014wide,
  title={Wide bandwidth and high coupling efficiency Si3N4-on-SOI dual-level grating coupler},
  author={Sacher, Wesley D and Huang, Ying and Ding, Liang and Taylor, Benjamin JF and Jayatilleka, Hasitha and Lo, Guo-Qiang and Poon, Joyce KS},
  journal={Optics express},
  volume={22},
  number={9},
  pages={10938--10947},
  year={2014},
  publisher={Optical Society of America}
}

@article{ong2018sinx,
  title={SiNx bilayer grating coupler for photonic systems},
  author={Ong, Eng Wen and Fahrenkopf, Nicholas M and Coolbaugh, Douglas D},
  journal={OSA Continuum},
  volume={1},
  number={1},
  pages={13--25},
  year={2018},
  publisher={Optical Society of America}
}

@article{romero2013visible,
  title={Visible wavelength silicon nitride focusing grating coupler with AlCu/TiN reflector},
  author={Romero-Garc{\'\i}a, Sebastian and Merget, Florian and Zhong, Frank and Finkelstein, Hod and Witzens, Jeremy},
  journal={Optics letters},
  volume={38},
  number={14},
  pages={2521--2523},
  year={2013},
  publisher={Optical Society of America}
}

@article{hong2019high,
  title={A high efficiency silicon nitride waveguide grating coupler with a multilayer bottom reflector},
  author={Hong, Jianxun and Spring, Andrew M and Qiu, Feng and Yokoyama, Shiyoshi},
  journal={Scientific reports},
  volume={9},
  number={1},
  pages={12988},
  year={2019},
  publisher={Nature Publishing Group UK London}
}

@article{ruiz2025inverse,
  title={Inverse-designed 90-degree silicon nitride bends for the C band},
  author={Ruiz, Julian L Pita and Dalvand, Narges and M{\'e}nard, Micha{\"e}l},
  journal={Journal of Lightwave Technology},
  year={2025},
  publisher={IEEE}
}

@article{nambiar2019high,
  title={High efficiency DBR assisted grating chirp generators for silicon nitride fiber-chip coupling},
  author={Nambiar, Siddharth and Ranganath, Praveen and Kallega, Rakshitha and Selvaraja, Shankar Kumar},
  journal={Scientific reports},
  volume={9},
  number={1},
  pages={18821},
  year={2019},
  publisher={Nature Publishing Group UK London}
}

@article{korvcek2023low,
  title={Low-loss grating coupler based on inter-layer mode interference in a hybrid silicon nitride platform},
  author={Kor{\v{c}}ek, Radovan and Cheben, Pavel and Fraser, William and Schmid, Jens H and Milanizadeh, Maziyar and Alonso-Ramos, Carlos and Ye, Winnie N and Benedikovi{\v{c}}, Daniel},
  journal={Optics Letters},
  volume={48},
  number={15},
  pages={4017--4020},
  year={2023},
  publisher={Optica Publishing Group}
}

@article{pita2025integrated,
  title={Integrated silicon nitride devices via inverse design},
  author={Pita Ruiz, Julian L and Dalvand, Narges and M{\'e}nard, Micha{\"e}l},
  journal={Nature Communications},
  volume={16},
  number={1},
  pages={9307},
  year={2025},
  publisher={Nature Publishing Group UK London}
}

@inproceedings{baets2016silicon,
  title={Silicon photonics: Silicon nitride versus silicon-on-insulator},
  author={Baets, Roel and Subramanian, Ananth Z and Clemmen, St{\'e}phane and Kuyken, Bart and Bienstman, Peter and Le Thomas, Nicolas and Roelkens, G{\"u}nther and Van Thourhout, Dries and Helin, Philippe and Severi, Simone},
  booktitle={Optical fiber communication conference},
  pages={Th3J--1},
  year={2016},
  organization={Optica Publishing Group}
}

@article{zhao2020design,
  title={Design principles of apodized grating couplers},
  author={Zhao, Zhexin and Fan, Shanhui},
  journal={Journal of Lightwave Technology},
  volume={38},
  number={16},
  pages={4435--4446},
  year={2020},
  publisher={OSA}
}

@article{wang2019robust,
  title={Robust design of topology-optimized metasurfaces},
  author={Wang, Evan W and Sell, David and Phan, Thaibao and Fan, Jonathan A},
  journal={Optical Materials Express},
  volume={9},
  number={2},
  pages={469--482},
  year={2019},
  publisher={Optical Society of America}
}

@article{sapra2019inverse,
  title={Inverse design and demonstration of broadband grating couplers},
  author={Sapra, Neil V and Vercruysse, Dries and Su, Logan and Yang, Ki Youl and Skarda, Jinhie and Piggott, Alexander Y and Vu{\v{c}}kovi{\'c}, Jelena},
  journal={IEEE Journal of Selected Topics in Quantum Electronics},
  volume={25},
  number={3},
  pages={1--7},
  year={2019},
  publisher={IEEE}
}

@article{den2016optical,
  title={Optical wafer metrology sensors for process-robust CD and overlay control in semiconductor device manufacturing},
  author={Den Boef, Arie J},
  journal={Surface Topography: Metrology and Properties},
  volume={4},
  number={2},
  pages={023001},
  year={2016},
  publisher={IOP Publishing}
}

@article{michaels2018leveraging,
  title={Leveraging continuous material averaging for inverse electromagnetic design},
  author={Michaels, Andrew and Yablonovitch, Eli},
  journal={Optics express},
  volume={26},
  number={24},
  pages={31717--31737},
  year={2018},
  publisher={Optical Society of America}
}

@article{orji2018metrology,
  title={Metrology for the next generation of semiconductor devices},
  author={Orji, Ndubuisi G and Badaroglu, Mustafa and Barnes, Bryan M and Beitia, Carlos and Bunday, Benjamin D and Celano, Umberto and Kline, Regis J and Neisser, Mark and Obeng, Yaw and Vladar, AE},
  journal={Nature electronics},
  volume={1},
  number={10},
  pages={532--547},
  year={2018},
  publisher={Nature Publishing Group UK London}
}

@article{mehta2016integrated,
  title={Integrated optical addressing of an ion qubit},
  author={Mehta, Karan K and Bruzewicz, Colin D and McConnell, Robert and Ram, Rajeev J and Sage, Jeremy M and Chiaverini, John},
  journal={Nature nanotechnology},
  volume={11},
  number={12},
  pages={1066--1070},
  year={2016},
  publisher={Nature Publishing Group UK London}
}

@article{bruzewicz2019trapped,
  title={Trapped-ion quantum computing: Progress and challenges},
  author={Bruzewicz, Colin D and Chiaverini, John and McConnell, Robert and Sage, Jeremy M},
  journal={Applied physics reviews},
  volume={6},
  number={2},
  year={2019},
  publisher={AIP Publishing}
}

@book{mack2008fundamental,
  title={Fundamental principles of optical lithography: the science of microfabrication},
  author={Mack, Chris},
  year={2008},
  publisher={John Wiley \& Sons}
}

@book{bendsoe2013topology,
  title={Topology optimization: theory, methods, and applications},
  author={Bendsoe, Martin Philip and Sigmund, Ole},
  year={2013},
  publisher={Springer Science \& Business Media}
}

@article{bendsoe1988generating,
  title={Generating optimal topologies in structural design using a homogenization method},
  author={Bends{\o}e, Martin Philip and Kikuchi, Noboru},
  journal={Computer methods in applied mechanics and engineering},
  volume={71},
  number={2},
  pages={197--224},
  year={1988},
  publisher={Elsevier}
}

@article{gertler2025many,
  title={Many photonic design problems are sparse QCQPs},
  author={Gertler, Shai and Kuang, Zeyu and Christie, Colin and Li, Hao and Miller, Owen D},
  journal={Science Advances},
  volume={11},
  number={1},
  pages={eadl3237},
  year={2025},
  publisher={American Association for the Advancement of Science}
}

@article{kuang2020computational,
  title={Computational bounds to light--matter interactions via local conservation laws},
  author={Kuang, Zeyu and Miller, Owen D},
  journal={Physical Review Letters},
  volume={125},
  number={26},
  pages={263607},
  year={2020},
  publisher={APS}
}

@article{ma2021deep,
  title={Deep learning for the design of photonic structures},
  author={Ma, Wei and Liu, Zhaocheng and Kudyshev, Zhaxylyk A and Boltasseva, Alexandra and Cai, Wenshan and Liu, Yongmin},
  journal={Nature photonics},
  volume={15},
  number={2},
  pages={77--90},
  year={2021},
  publisher={Nature Publishing Group UK London}
}

@article{bennet2024illustrated,
  title={Illustrated tutorial on global optimization in nanophotonics},
  author={Bennet, Pauline and Langevin, Denis and Essoual, Chaymae and Khaireh-Walieh, Abdourahman and Teytaud, Olivier and Wiecha, Peter and Moreau, Antoine},
  journal={Journal of the Optical Society of America B},
  volume={41},
  number={2},
  pages={A126--A145},
  year={2024},
  publisher={Optica Publishing Group}
}

@article{lin2021end,
  title={End-to-end nanophotonic inverse design for imaging and polarimetry},
  author={Lin, Zin and Roques-Carmes, Charles and Pestourie, Rapha{\"e}l and Solja{\v{c}}i{\'c}, Marin and Majumdar, Arka and Johnson, Steven G},
  journal={Nanophotonics},
  volume={10},
  number={3},
  pages={1177--1187},
  year={2021},
  publisher={De Gruyter}
}

@article{yang2020inverse,
  title={Inverse-designed non-reciprocal pulse router for chip-based LiDAR},
  author={Yang, Ki Youl and Skarda, Jinhie and Cotrufo, Michele and Dutt, Avik and Ahn, Geun Ho and Sawaby, Mahmoud and Vercruysse, Dries and Arbabian, Amin and Fan, Shanhui and Al{\`u}, Andrea and others},
  journal={Nature Photonics},
  volume={14},
  number={6},
  pages={369--374},
  year={2020},
  publisher={Nature Publishing Group UK London}
}

@article{yang2022multi,
  title={Multi-dimensional data transmission using inverse-designed silicon photonics and microcombs},
  author={Yang, Ki Youl and Shirpurkar, Chinmay and White, Alexander D and Zang, Jizhao and Chang, Lin and Ashtiani, Farshid and Guidry, Melissa A and Lukin, Daniil M and Pericherla, Srinivas V and Yang, Joshua and others},
  journal={Nature communications},
  volume={13},
  number={1},
  pages={7862},
  year={2022},
  publisher={Nature Publishing Group UK London}
}

@article{ganapati2013light,
  title={Light trapping textures designed by electromagnetic optimization for subwavelength thick solar cells},
  author={Ganapati, Vidya and Miller, Owen D and Yablonovitch, Eli},
  journal={IEEE Journal of Photovoltaics},
  volume={4},
  number={1},
  pages={175--182},
  year={2013},
  publisher={IEEE}
}

@article{li2022inverse,
  title={Inverse design enables large-scale high-performance meta-optics reshaping virtual reality},
  author={Li, Zhaoyi and Pestourie, Rapha{\"e}l and Park, Joon-Suh and Huang, Yao-Wei and Johnson, Steven G and Capasso, Federico},
  journal={Nature communications},
  volume={13},
  number={1},
  pages={2409},
  year={2022},
  publisher={Nature Publishing Group UK London}
}

@article{wang2003level,
  title={A level set method for structural topology optimization},
  author={Wang, Michael Yu and Wang, Xiaoming and Guo, Dongming},
  journal={Computer methods in applied mechanics and engineering},
  volume={192},
  number={1-2},
  pages={227--246},
  year={2003},
  publisher={Elsevier}
}

@article{van2013level,
  title={Level-set methods for structural topology optimization: a review},
  author={Van Dijk, Nico P and Maute, Kurt and Langelaar, Matthijs and Van Keulen, Fred},
  journal={Structural and Multidisciplinary Optimization},
  volume={48},
  number={3},
  pages={437--472},
  year={2013},
  publisher={Springer}
}

@article{lalau2013adjoint,
  title={Adjoint shape optimization applied to electromagnetic design},
  author={Lalau-Keraly, Christopher M and Bhargava, Samarth and Miller, Owen D and Yablonovitch, Eli},
  journal={Optics express},
  volume={21},
  number={18},
  pages={21693--21701},
  year={2013},
  publisher={Optical Society of America}
}

@article{jensen2005topology,
  title={Topology optimization of photonic crystal structures: a high-bandwidth low-loss T-junction waveguide},
  author={Jensen, Jakob S and Sigmund, Ole},
  journal={Journal of the Optical Society of America B},
  volume={22},
  number={6},
  pages={1191--1198},
  year={2005},
  publisher={Optical Society of America}
}

@article{eschenauer1994bubble,
  title={Bubble method for topology and shape optimization of structures},
  author={Eschenauer, Hans A and Kobelev, Vladimir V and Schumacher, Axel},
  journal={Structural optimization},
  volume={8},
  number={1},
  pages={42--51},
  year={1994},
  publisher={Springer}
}

@article{amstutz2006new,
  title={A new algorithm for topology optimization using a level-set method},
  author={Amstutz, Samuel and Andr{\"a}, Heiko},
  journal={Journal of computational physics},
  volume={216},
  number={2},
  pages={573--588},
  year={2006},
  publisher={Elsevier}
}

@article{allaire2005structural,
  title={Structural optimization using topological and shape sensitivity via a level set method},
  author={Allaire, Gr{\'e}goire and Gournay, F de and Jouve, Fran{\c{c}}ois and Toader, A-M},
  journal={Control and cybernetics},
  volume={34},
  number={1},
  pages={59--80},
  year={2005},
  publisher={Polska Akademia Nauk. Instytut Bada{\'n} Systemowych PAN}
}

@article{hughes2021perspective,
  title={A perspective on the pathway toward full wave simulation of large area metalenses},
  author={Hughes, Tyler W and Minkov, Momchil and Liu, Victor and Yu, Zongfu and Fan, Shanhui},
  journal={Applied Physics Letters},
  volume={119},
  number={15},
  year={2021},
  publisher={AIP Publishing}
}

@misc{tidy3d,
  title  = {{Tidy3D} electromagnetic solver},
  author = {Flexcompute, Inc.},
  year   = {2024},
  note   = {\url{https://www.flexcompute.com/tidy3d/}},
}

@article{tamir1977analysis,
  title={Analysis and design of grating couplers},
  author={Tamir, Theodor and Peng, Song-Tsuen},
  journal={Applied physics},
  volume={14},
  number={3},
  pages={235--254},
  year={1977},
  publisher={Springer}
}

@incollection{petit1980tutorial,
  title={A tutorial introduction},
  author={Petit, R},
  booktitle={Electromagnetic Theory of Gratings},
  pages={1--52},
  year={1980},
  publisher={Springer}
}

@article{chen2010apodized,
  title={Apodized waveguide grating couplers for efficient coupling to optical fibers},
  author={Chen, Xia and Li, Chao and Fung, Christy KY and Lo, Stanley MG and Tsang, Hon K},
  journal={IEEE Photonics Technology Letters},
  volume={22},
  number={15},
  pages={1156--1158},
  year={2010},
  publisher={IEEE}
}

@article{marchetti2017high,
  title={High-efficiency grating-couplers: demonstration of a new design strategy},
  author={Marchetti, Riccardo and Lacava, Cosimo and Khokhar, Ali and Chen, Xia and Cristiani, Ilaria and Richardson, David J and Reed, Graham T and Petropoulos, Periklis and Minzioni, Paolo},
  journal={Scientific reports},
  volume={7},
  number={1},
  pages={16670},
  year={2017},
  publisher={Nature Publishing Group UK London}
}

@article{wang1990guided,
  title={Guided-mode resonances in planar dielectric-layer diffraction gratings},
  author={Wang, SS and Magnusson, Robert and Bagby, Jonathan S and Moharam, MG},
  journal={Journal of the Optical Society of America A},
  volume={7},
  number={8},
  pages={1470--1474},
  year={1990},
  publisher={Optical Society of America}
}

@article{rosenblatt1997resonant,
  title={Resonant grating waveguide structures},
  author={Rosenblatt, David and Sharon, Avener and Friesem, Asher A},
  journal={IEEE Journal of Quantum electronics},
  volume={33},
  number={11},
  pages={2038--2059},
  year={1997},
  publisher={IEEE}
}

@article{quaranta2018recent,
  title={Recent advances in resonant waveguide gratings},
  author={Quaranta, Giorgio and Basset, Guillaume and Martin, Olivier JF and Gallinet, Benjamin},
  journal={Laser \& Photonics Reviews},
  volume={12},
  number={9},
  pages={1800017},
  year={2018},
  publisher={Wiley Online Library}
}

@article{tang2010highly,
  title={Highly efficient nonuniform grating coupler for silicon-on-insulator nanophotonic circuits},
  author={Tang, Yongbo and Wang, Zhechao and Wosinski, Lech and Westergren, Urban and He, Sailing},
  journal={Optics letters},
  volume={35},
  number={8},
  pages={1290--1292},
  year={2010},
  publisher={Optical Society of America}
}

@article{taillaert2002out,
  title={An out-of-plane grating coupler for efficient butt-coupling between compact planar waveguides and single-mode fibers},
  author={Taillaert, Dirk and Bogaerts, Wim and Bienstman, Peter and Krauss, Thomas F and Van Daele, Peter and Moerman, Ingrid and Verstuyft, Steven and De Mesel, Kurt and Baets, Roel},
  journal={IEEE Journal of Quantum Electronics},
  volume={38},
  number={7},
  pages={949--955},
  year={2002},
  publisher={IEEE}
}

@article{taillaert2006grating,
  title={Grating couplers for coupling between optical fibers and nanophotonic waveguides},
  author={Taillaert, Dirk and Laere, Frederik Van and Ayre, Melanie and Bogaerts, Wim and Thourhout, Dries Van and Bienstman, Peter and Baets, Roel},
  journal={Japanese Journal of Applied Physics},
  volume={45},
  number={8R},
  pages={6071},
  year={2006}
}

@article{ding2014fully,
  title={Fully etched apodized grating coupler on the SOI platform with- 0.58 dB coupling efficiency},
  author={Ding, Yunhong and Peucheret, Christophe and Ou, Haiyan and Yvind, Kresten},
  journal={Optics letters},
  volume={39},
  number={18},
  pages={5348--5350},
  year={2014},
  publisher={Optical Society of America}
}

@article{zaoui2014bridging,
  title={Bridging the gap between optical fibers and silicon photonic integrated circuits},
  author={Zaoui, Wissem Sfar and Kunze, Andreas and Vogel, Wolfgang and Berroth, Manfred and Butschke, J{\"o}rg and Letzkus, Florian and Burghartz, Joachim},
  journal={Optics express},
  volume={22},
  number={2},
  pages={1277--1286},
  year={2014},
  publisher={Optical Society of America}
}

@article{hammond2022multi,
  title={Multi-layer inverse design of vertical grating couplers for high-density, commercial foundry interconnects},
  author={Hammond, Alec M and Slaby, Joel B and Probst, Michael J and Ralph, Stephen E},
  journal={Optics Express},
  volume={30},
  number={17},
  pages={31058--31072},
  year={2022},
  publisher={Optica Publishing Group}
}

@article{su2018fully,
  title={Fully-automated optimization of grating couplers},
  author={Su, Logan and Trivedi, Rahul and Sapra, Neil V and Piggott, Alexander Y and Vercruysse, Dries and Vu{\v{c}}kovi{\'c}, Jelena},
  journal={Optics express},
  volume={26},
  number={4},
  pages={4023--4034},
  year={2018},
  publisher={Optical Society of America}
}

@article{van2021wafer,
  title={Wafer-level testing of inverse-designed and adjoint-inspired vertical grating coupler designs compatible with DUV lithography},
  author={Van Vaerenbergh, Thomas and Sun, Peng and Hooten, Sean and Jain, Mudit and Wilmart, Quentin and Seyedi, Ashkan and Huang, Zhihong and Fiorentino, Marco and Beausoleil, Ray},
  journal={Optics Express},
  volume={29},
  number={23},
  pages={37021--37036},
  year={2021},
  publisher={Optical Society of America}
}

@article{van2022wafer,
  title={Wafer-level testing of inverse-designed and adjoint-inspired dual layer Si-SiN vertical grating couplers},
  author={Van Vaerenbergh, Thomas and Hooten, Sean and Jain, Mudit and Sun, Peng and Wilmart, Quentin and Seyedi, Ashkan and Huang, Zhihong and Fiorentino, Marco and Beausoleil, Ray},
  journal={Journal of Physics: Photonics},
  volume={4},
  number={4},
  pages={044001},
  year={2022},
  publisher={IOP Publishing}
}

@article{taillaert2003compact,
  title={A compact two-dimensional grating coupler used as a polarization splitter},
  author={Taillaert, Dirk and Chong, Harold and Borel, Peter I and Frandsen, Lars H and De La Rue, Richard M and Baets, Roel},
  journal={IEEE Photonics Technology Letters},
  volume={15},
  number={9},
  pages={1249--1251},
  year={2003},
  publisher={IEEE}
}

@article{piggott2014inverse,
  title={Inverse design and implementation of a wavelength demultiplexing grating coupler},
  author={Piggott, Alexander Y and Lu, Jesse and Babinec, Thomas M and Lagoudakis, Konstantinos G and Petykiewicz, Jan and Vu{\v{c}}kovi{\'c}, Jelena},
  journal={Scientific reports},
  volume={4},
  number={1},
  pages={7210},
  year={2014},
  publisher={Nature Publishing Group UK London}
}

@article{sideris2022foundry,
  title={Foundry-fabricated grating coupler demultiplexer inverse-designed via fast integral methods},
  author={Sideris, Constantine and Khachaturian, Aroutin and White, Alexander D and Bruno, Oscar P and Hajimiri, Ali},
  journal={Communications Physics},
  volume={5},
  number={1},
  pages={68},
  year={2022},
  publisher={Nature Publishing Group UK London}
}

@article{streshinsky2013compact,
  title={A compact bi-wavelength polarization splitting grating coupler fabricated in a 220 nm SOI platform},
  author={Streshinsky, Matthew and Shi, Ruizhi and Novack, Ari and Cher, Roger Tern Poh and Lim, Andy Eu-Jin and Lo, Patrick Guo-Qiang and Baehr-Jones, Tom and Hochberg, Michael},
  journal={Optics express},
  volume={21},
  number={25},
  pages={31019--31028},
  year={2013},
  publisher={Optical Society of America}
}

@article{cheng2022single,
  title={Single-step etched two-dimensional polarization splitting dual-band grating coupler for wavelength (de) multiplexing},
  author={Cheng, Guanglian and Yi, Qiyuan and Li, Qiyuan and Yan, Zhiwei and Xu, Fanglu and Zou, Yongchao and Li, Ting and Zou, Yi and Yu, Yu and Shen, Li},
  journal={Optics Letters},
  volume={47},
  number={15},
  pages={3924--3927},
  year={2022},
  publisher={Optica Publishing Group}
}

@article{su2024topology,
  title={Topology optimization enables high-Q metasurface for color selectivity},
  author={Su, Huan-Teng and Wang, Lu-Yun and Hsu, Chih-Yao and Wu, Yun-Chien and Lin, Chang-Yi and Chang, Shu-Ming and Huang, Yao-Wei},
  journal={Nano Letters},
  volume={24},
  number={33},
  pages={10055--10061},
  year={2024},
  publisher={ACS Publications}
}

@article{roelkens2006high,
  title={High efficiency Silicon-on-Insulator grating coupler based on a poly-Silicon overlay},
  author={Roelkens, G{\"u}nther and Van Thourhout, Dries and Baets, Roel},
  journal={Optics Express},
  volume={14},
  number={24},
  pages={11622--11630},
  year={2006},
  publisher={Optical Society of America}
}

@article{vermeulen2010high,
  title={High-efficiency fiber-to-chip grating couplers realized using an advanced CMOS-compatible silicon-on-insulator platform},
  author={Vermeulen, Diedrik and Selvaraja, Shankar and Verheyen, Peter and Lepage, Guy and Bogaerts, Wim and Absil, Philippe and Van Thourhout, Dries and Roelkens, Gunther},
  journal={Optics express},
  volume={18},
  number={17},
  pages={18278--18283},
  year={2010},
  publisher={Optical Society of America}
}

@article{zhu2022optical,
  title={Optical wafer defect inspection at the 10 nm technology node and beyond},
  author={Zhu, Jinlong and Liu, Jiamin and Xu, Tianlai and Yuan, Shuai and Zhang, Zexu and Jiang, Hao and Gu, Honggang and Zhou, Renjie and Liu, Shiyuan},
  journal={International Journal of Extreme Manufacturing},
  volume={4},
  number={3},
  pages={032001},
  year={2022},
  publisher={IOP Publishing}
}

@article{vieu2000electron,
  title={Electron beam lithography: resolution limits and applications},
  author={Vieu, Christophe and Carcenac, F and P{\'e}pin, Anne and Chen, Yong and Mejias, Marcelo and Lebib, Amira and Manin-Ferlazzo, L and Couraud, L and Launois, H},
  journal={Applied surface science},
  volume={164},
  number={1-4},
  pages={111--117},
  year={2000},
  publisher={Elsevier}
}

@article{dai2012passive,
  title={Passive technologies for future large-scale photonic integrated circuits on silicon: polarization handling, light non-reciprocity and loss reduction},
  author={Dai, Daoxin and Bauters, Jared and Bowers, John E},
  journal={Light: Science \& Applications},
  volume={1},
  number={3},
  pages={e1--e1},
  year={2012},
  publisher={Nature Publishing Group}
}

@article{shi2020scaling,
  title={Scaling capacity of fiber-optic transmission systems via silicon photonics},
  author={Shi, Wei and Tian, Ye and Gervais, Antoine},
  journal={Nanophotonics},
  volume={9},
  number={16},
  pages={4629--4663},
  year={2020},
  publisher={De Gruyter}
}

@article{mekis2010grating,
  title={A grating-coupler-enabled {CMOS} photonics platform},
  author={Mekis, Attila and Gloeckner, Steffen and Masini, Gianlorenzo and Narasimha, Adithyaram and Pinguet, Thierry and Sahni, Subal and De Dobbelaere, Peter},
  journal={IEEE Journal of Selected Topics in Quantum Electronics},
  volume={17},
  number={3},
  pages={597--608},
  year={2010},
  publisher={IEEE}
}

@article{blumenthal2024enabling,
  title={Enabling photonic integrated 3D magneto-optical traps for quantum sciences and applications},
  author={Blumenthal, Daniel J and Isichenko, Andrei and Chauhan, Nitesh},
  journal={Optica Quantum},
  volume={2},
  number={6},
  pages={444--457},
  year={2024},
  publisher={Optica Publishing Group}
}

@article{marchetti2019coupling,
  title={Coupling strategies for silicon photonics integrated chips},
  author={Marchetti, Riccardo and Lacava, Cosimo and Carroll, Lee and Gradkowski, Kamil and Minzioni, Paolo},
  journal={Photonics Research},
  volume={7},
  number={2},
  pages={201--239},
  year={2019},
  publisher={Chinese Laser Press and Optical Society of America}
}

@article{sun2015single,
  title={Single-chip microprocessor that communicates directly using light},
  author={Sun, Chen and Wade, Mark T and Lee, Yunsup and Orcutt, Jason S and Alloatti, Luca and Georgas, Michael S and Waterman, Andrew S and Shainline, Jeffrey M and Avizienis, Rimas R and Lin, Sen and others},
  journal={Nature},
  volume={528},
  number={7583},
  pages={534--538},
  year={2015},
  publisher={Nature Publishing Group UK London}
}

@article{niffenegger2020integrated,
  title={Integrated multi-wavelength control of an ion qubit},
  author={Niffenegger, Robert J and Stuart, Jules and Sorace-Agaskar, Cheryl and Kharas, Dave and Bramhavar, Suraj and Bruzewicz, Colin D and Loh, William and Maxson, Ryan T and McConnell, Robert and Reens, David and others},
  journal={Nature},
  volume={586},
  number={7830},
  pages={538--542},
  year={2020},
  publisher={Nature Publishing Group UK London}
}

@article{mehta2020integrated,
  title={Integrated optical multi-ion quantum logic},
  author={Mehta, Karan K and Zhang, Chi and Malinowski, Maciej and Nguyen, Thanh-Long and Stadler, Martin and Home, Jonathan P},
  journal={Nature},
  volume={586},
  number={7830},
  pages={533--537},
  year={2020},
  publisher={Nature Publishing Group UK London}
}


\end{document}